\documentclass[twocolumn]{article}
\usepackage{graphicx}
\usepackage{amsmath}
\usepackage{amssymb}
\begin{document}
\title{Optical chaos in nonlinear photonic crystals}
\author{
Kirill N. Alekseev$^{1,2,3}$\thanks{E-mail: Kirill.Alekseev@oulu.fi},
Aleksey V. Ponomarev$^{3,4}$\\
$^{1}$Department of Physical Sciences, P.O. Box 3000, University of Oulu FIN-90014,
Finland\\
$^{2}$Max-Planck-Institute f\"ur Physik komplexer Systeme,\\
N\"{o}thnitzer Str. 38, D-01187 Dresden, Germany\\
$^{3}$Theory of Nonlinear Processes Laboratory, Kirensky Institute of
Physics,\\
Russian Academy of Sciences, Krasnoyarsk 660036, Russia\\
$^{4}$ Department of Physics, Krasnoyarsk State University,
Krasnoyarsk 660041, Russia}
\date{}
\maketitle
\begin{abstract}
We examine a spatial evolution of lightwaves in a nonlinear photonic
crystal with a quadratic nonlinearity when simultaneously
a second harmonic and a sum-frequency generation are quasi-phase-matched.
We find the conditions of a transition to Hamiltonian chaos
for different amplitudes of lightwaves at the boundary of the crystal.
\end{abstract}

\medskip
\hrule
\bigskip


Wave mixing in nonlinear optical materials is a basis of modern optical
sciences and technologies.
Cascading several wave-mixing processes in the same low-loss material
one can in principle achieve a high efficiency using a large value of the
lowest-order optical nonlinearity.
The theoretical investigations of cascading of several scalar optical
three-wave-mixing processes
in the bulk materials with ${\chi}^{(2)}$ nonlinearity has a long history
\cite{akhmanov}. In particular, Akhmanov and co-workers have found the
efficiency of a third harmonic generation (THG) via cascading of a second
harmonic generation (SHG) and a sum-frequency mixing (SFM) in a quadratic
medium \cite{akhmanov1}, while Komissarova and Sukhorukov have described an
efficient parametric amplification at a high-frequency pump in the same system
\cite{komissarova}.
Obviously, the observation of these nonlinear effects demands a simultaneous
satisfaction of phase-matching conditions for several parametric processes
as perfectly as possible. On other hand, it has been shown later that the
systems,
for which several optical wave-mixing processes can be simultaneously phase-matched,
are in general nonintegrable; therefore
a competition of two (or more) parametric processes can often result
in a chaotic spatial evolution of lightwaves \cite{alekseev,alekseeva}.
However, until nowadays it was unclear how to achieve a phase-matching for
several processes in {\it homogeneous medium}
employing traditional techniques, such as using a birefringence in
ferroelectric crystals.
\par
The solution of this problem has been found rather recently
\cite{aleksandrovski,pfister,sl-theor}; it consists in an introduction of the
different types of {\it artificial periodicity} of a nonlinear medium
resulting in a formation of nonlinear 1D
and 2D superstructures termed {\em optical superlattices} \cite{china-review}
or {\em nonlinear photonic crystals} (NPCs) \cite{2D-photon-cr}. In NPCs there
is a periodic (or quasiperiodic) spatial variation of a nonlinear
susceptibility tensor while a linear susceptibility tensor is constant.
\par
In these engineering nonlinear materials a phase mismatch
between the interacting lightwaves could be compensated by the Bragg vector
of NPC. The idea of such kind of {\em quasi-phase-matching} (QPM)
was introduced by Bloembergen and co-workers many years ago \cite{armstrong}.
However, only recently the rapid progress in a fabrication of high quality
ferroelectric crystals with a periodic domain inversion made the QPM method
very popular \cite{china-review,chirkin-review}.
We should stress that the conditions for QPM may be
fulfilled for several wave-mixing processes simultaneously;
the QPM also has an advantage as using of largest nonlinear coefficient.
\par
Nowadays there are several experiments on an
observation of third and fourth harmonics in different periodically or
quasiperiodically poled
ferroelectric crystals with ${\chi}^{(2)}$ nonlinearity
\cite{pfister,volkov,thg-exp}, which
clearly demonstrate an importance of multiple-mixing in NPCs for
the potential applications.
Modern theoretical activities on the nonlinear lightwaves interactions in NPCs
are mainly focused on the studies of strong energy interchange between the
waves \cite{chirkin-review} (this is a development of the earlier activities
\cite{akhmanov1,komissarova}),
as well as on the formation of spatial optical solitons \cite{solitons}.
\par
In this work we describe a novel for the physics of NPCs
effect of Hamiltonian optical chaos. Namely, we show that spatial evolution of
three light waves, participating simultaneously in
SHG and SFM in the conditions of QPM, is chaotic for many values of complex
amplitude of the waves at the boundary of ${\chi}^{(2)}$-NPC.
There also exists an integrable limit when the evolution of waves
is always regular regardless of absolute values of their  complex amplitudes.
The integrable limit corresponds to the particular values of two combinations
of wave phases at the boundary of nonlinear medium.
In particular, the problem of THG belongs to the integrable limit, therefore
in the conditions of recent experiments \cite{pfister,volkov,thg-exp},
nonlinear light dynamics should be always regular. However, even a rather small
change in amplitudes and phases of waves at the boundary of crystal,
in respect to those considered in \cite{pfister,volkov,thg-exp},
should result in a transition to chaos.
\par
We consider a spatial evolution of three co-propagating plane waves
$$
E=\frac{1}{2}\sum_{j=1}^3 A_j \exp{[i(j\omega t - k_j z)]}+
\mbox{c.c.},\quad k_j=k(j\omega)
$$
in a periodically poled crystal in the conditions when simultaneously SHG,
$\omega+\omega\rightarrow2\omega$, and SFM,
$\omega+2\omega\rightarrow 3\omega$ take place.
Equations of motion for the slowly varying complex amplitudes
$A_l$ ($l=1,2,3)$ of
the waves are \cite{chirkin-review,china-review}
\begin{eqnarray}
\label{A-motion}
\frac{d A_1}{d z}&=&-i\beta_3 g(z) A_3 A_2^* e^{-i \Delta k_3 z}
-i\beta_2 g(z) A_2 A_1^* e^{-i \Delta k_2 z},\nonumber\\
\frac{d A_2}{d z}&=&-i 2 \beta_3 g(z) A_3 A_1^* e^{-i \Delta k_3 z}
-i\beta_2 g(z) A_1^2 e^{i \Delta k_2 z},\\
\frac{d A_3}{d z}&=&-i 3 \beta_3 g(z) A_1 A_2 e^{i \Delta k_3 z},\nonumber
\end{eqnarray}
where $g(z)$ is a function that equals to +1 (or -1) in a single positive
(negative) polarization domain of the ferroelectric crystal.
In this work for the sake of simplicity, we consider only a periodic alternative
domain superlattice with a spatial period $\Lambda$. However, $g(z)$ can be
a quasiperiodic function in the case of nonlinear quasicrystals
\cite{sl-theor,china-review}. Note that we consider a typical situation
$\lambda\ll\Lambda$, where $\lambda$ is a wavelength
\cite{china-review,chirkin-review,thg-exp}.
\par
The coupling constants between waves, $\beta_2$ and $\beta_3$, are defined as
$$
\beta_{2,3}=\frac{\omega d_{eff}}{c n},
$$
where $d_{eff}=2\pi{\chi}^{(2)}$ and $n$ is a refractive index. 
The refractive indexes for the different waves are different due to a light 
dispersion. However, it may be shown that $\Delta n/n\simeq\lambda/\Lambda
\ll 1$ in the conditions of QPM,
therefore in what follows we will take $\beta_2=\beta_3\equiv\beta$.
Finally, the phase mismatches involved in Eqs. (\ref{A-motion}) are
$\Delta k_2= k_2-2 k_1$ and $\Delta k_3= k_3-k_2-k_1$.
Let both these mismatches be compensated by a reciprocal lattice vector of
NPC, that is
\begin{equation}
\label{QPM-def}
\Delta k_2=2\pi m_1/\Lambda, \quad\Delta k_3=2\pi m_2/\Lambda,
\end{equation}
where $m_j=\pm 1,\pm 3,\pm 5, \ldots$. The methods of achievement of QPM for
several parametric processes in a single NPC have been recently discussed in
Refs.~\cite{aleksandrovski,sl-theor,2D-photon-cr} (theory) and
Refs.~\cite{pfister,volkov,thg-exp} (experiment).
\par
The dynamical system (\ref{A-motion}) together with initial conditions,
which in
our case are the values of complex amplitudes at the boundary of NPC,
$A_j(z=0)$, completely determine the nonlinear spatial evolution of waves.
Before specification of these initial conditions, we can further simplify the
equations of motion. First, we introduce new scaled amplitudes
$a_l=A_l/(\sqrt{l} A_0)$, where $l=1,2,3$ and
$A_0\equiv\max{\left( |A_1(0)|,|A_2(0)|,|A_3(0)|\right)}$.
Second, we make the Fourier expansion of the function $g(z)$
$$
g(z)=\sum_{n=1}^{\infty} \frac{4}{\pi n} \sin\left(\frac{2\pi n z}{\Lambda}
\right),
$$
where index $n$ takes only odd values. Now we substitute this
expansion into Eqs. (\ref{A-motion}), take into account the QPM conditions
(\ref{QPM-def}) and make an averaging of resulting motion equations over
``the short characteristic spatial scale'' $2\pi/\Lambda$. We have the
following basic equations
\begin{eqnarray}
\label{a-motion}
\dot{a}_1&=&-a_2 a_1^* - \xi a_3 a_2^*,\nonumber\\
\dot{a}_2&=&0.5 a_1^2 - \xi a_3 a_1^*,\\
\dot{a}_3&=&\xi a_1 a_2,\nonumber
\end{eqnarray}
where $\xi=\sqrt{3} m_2/m_3$ ($m_j$ are the quasi-phase matching orders, see
Eq.~(\ref{QPM-def}); we assume that $m_3\ge m_2$).
The overdots in Eqs. (\ref{a-motion}) mean the derivatives in respect to $z/l_{nl}$, and a characteristic nonlinear length $l_{nl}$ is defined as
\begin{equation}
\label{l_nl-def}
l_{nl}=\frac{\pi m_2}{2\sqrt{2}}\frac{1}{\beta A_0}.
\end{equation}
In the derivation of motion equations (\ref{a-motion}) we removed all fast
varying terms performing averaging over $2\pi/\Lambda$. It can be shown that
such a procedure is correct if $l_{nl}\gg\Lambda$ \cite{chirkin2}.
\par
The equations (\ref{a-motion}) can be presented in the canonical form with
the Hamiltonian function
\begin{eqnarray}
H &=&\left[ -i \left( \xi a_1^* a_2^* a_3+\frac{1}{2} a_1^{*2} a_2 \right)
\right]+\mbox{c.c.}, \label{Ham}\\
& &i\dot{a}_l=\frac{\partial H}{\partial a_l^*},\quad
i\dot{a}_l^*=-\frac{\partial H}{\partial a_l}.\nonumber
\end{eqnarray}
Additionally to the energy of wave interaction $E\equiv H$, (Eq.~(\ref{Ham})),
the dynamical system
(\ref{a-motion}) has the integral of motion
\begin{equation}
\label{int-motion}
|a_1|^2+2 |a_2|^2+3 |a_3|^2=1
\end{equation}
corresponding to the conservation of energy of noninteracting waves.
In a general case the system (\ref{a-motion}) does not have other global
integrals of motion, thus it is {\it nonintegrable} and
should demonstrate {\it chaotic dynamics} for many initial conditions
$a_l(0)$ \cite{ford,chaos-book}.
However, for some values of $\xi$ and some specific initial
conditions, an additional local integral of
motion can arise. Let us list these cases because they include the physically
important situations.
\par
First, if one of the parametric processes, either SHG or SFM, is
dominant ($\xi\ll 1$ or $\xi\gg 1$), then an additional integral of
motion arises, which is of the Manley-Rowe type \cite{chirkin-review}.
Second, nonlinear dynamics is strongly dependent on the initial values
of two ``resonant phases'' $\psi_2(0)$ and $\psi_3(0)$, where
\begin{equation}
\label{res-ph-def}
\psi_2=2\theta_1-\theta_2,\quad \psi_3=\theta_1+\theta_2-\theta_3
\end{equation}
and $\theta_j$ ($j=1,2,3$) are the lightwave phases, {\it i.e.}
$a_j=|a_j| \exp{(-i\theta_j)}$.
We found that {\it for $\psi_2(0)=\psi_3(0)=0$ dynamics is always regular}.
Moreover, using approaches of Refs.~\cite{komissarova,konotop00}, it is
possible to show that an additional local motion integral exists in this
case \cite{alekseev-unpub}. In particular, the problem of THG ($a_1(0)=1$,
$a_2(0)=a_3(0)=0$) belongs to this class of initial conditions. Therefore,
the spatial dynamics of lightwaves at THG is regular ({\it cf}
Ref.~\cite{zhang00}, where an analytic solution has been found).
\par
We have performed an intensive search of chaotic trajectories solving the equations of motion (\ref{a-motion}) numerically for two characteristic values of control parameter $\xi$ that correspond to the experimental situations described in works \cite{pfister} and \cite{volkov}, correspondingly:
\par
{\it ``Set I''}: The QPMs of first order for both processes, $m_1=m_3=1$,
$\xi=\sqrt{3}\approx 1.73$.
\par
{\it ``Set II''}: The QPMs of ninth and 33d orders, $m_1=9$, $m_3=33$, $\xi=
(3\sqrt{3})/11\approx 0.472$.
\par
We consider several types of initial conditions, which cover practically all
physically interesting cases (note that all these initial conditions satisfy
to the restriction arising from the integral of motion (\ref{int-motion})):
\par
{\it ``Problem 1''}: $a_1(0)=\alpha$, $a_2(0)=\left[ 1 -\alpha^2
\right]^{1/2} 2^{-1/2} \exp{(-i\phi)}$, $a_3(0)=0$,
where the real parameters $\phi$ and $\alpha$ vary in the ranges
$-\pi\leq\phi<\pi$
and $0\leq\alpha\leq 1$, correspondingly. Obviously here $|\psi_2(0)|=
|\psi_3(0)|=|\phi|$.
\par
{\it ``Problem 2''}: $a_1(0)=\left[ 1 -3\alpha^2\right]^{1/2} 3^{-1/2}
\exp{(-i\theta_1)}$,
$a_2(0)=\left[ 1 -3\alpha^2\right]^{1/2} 3^{-1/2}\exp{(-i\theta_2)}$,
$a_3(0)=\alpha\exp{(-i\theta_3)}$, where $-\pi\leq\theta_j<\pi$ ($j=1,2,3$)
and
$0\leq\alpha\leq 3^{-1/2}\approx 0.57735$.
\par
{\it ``Problem 3''}: $a_1(0)=\alpha\exp{(-i\theta_1)}$, $a_2(0)=0$,
$a_3(0)=\left[ 1 -\alpha^2\right]^{1/2} 3^{-1/2}\exp{(-i\theta_3)}$,
$-\pi\leq\theta_j<\pi$ ($j=1,3$) and $0\leq\alpha\leq 1$.
\par
We start our analysis with ``Problem 1''.
This set of initial conditions describes, in particular, the THG at $\alpha=1$
($\phi=0$) and the parametric amplification with a low-frequency pump at
$\alpha\ll 1$ \cite{chirkin-review}. In order to increase an efficiency of
energy transformation from a basic wave of frequency $\omega$ to a
wave of frequency $3\omega$, it was suggested recently to mix some nonzero
signal at the frequency $2\omega$ with a basic beam \cite{egorov}.
Such kind of initial conditions correspond to $\alpha\rightarrow 1$ (but $\alpha\neq 1$) with different values of phase $\phi$.
\par
\begin{figure}[htbp!]
\includegraphics[width=0.8\linewidth]{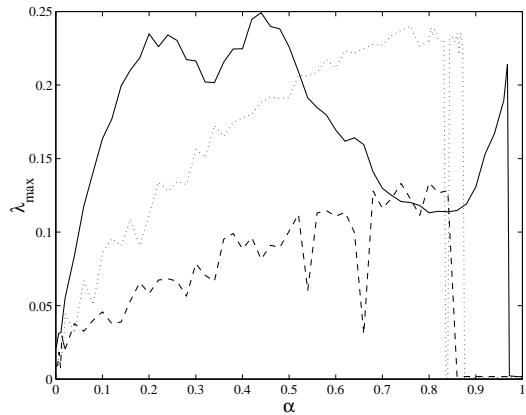}
\caption{Dependence of the value of maximal Lyapunov exponent
on the amplitude of first wave at the boundary of optical superlattice
$\alpha$ and for the different phases:  $\phi=-\pi/2$ (solid line),
$\phi=-0.1$ (dotted line) and $\phi=-0.01$ (dashed line). The first order
QPMs (problem 1, set I).}
\label{figps1}
\end{figure}

To distinguish regular and chaotic dynamics we compute the maximal Lyapunov
exponent $\lambda_{max}$ for the different values of initial
lightwave's amplitudes, $\alpha$,
and phases, $\phi$. For chaos  $\lambda_{max}>0$, in contrast
$\lambda_{max}=0$ for a regular motion \cite{chaos-book}.
The dependence of $\lambda_{max}$ on $\alpha$ for the first order
QPMs (set I) is depicted in Fig.~\ref{figps1}. For $\phi=0$ the initial
values of resonant phases,  $\psi_2(0)$ and $\psi_3(0)$, are zero
corresponding to
the integrable limit with $\lambda_{max}=0$ independently on the value of
$\alpha$ (not shown in Fig.~\ref{figps1}).
However, even a small deviation from the integrable limit, $|\psi_2(0)|=
|\psi_3(0)|=|\phi|=0.01$, results in a chaotic motion for a quite wide
range of initial conditions (dashed line).
Further increase in the value of $|\phi|$ makes chaos more strong (dotted
line, $|\phi|=0.1$); the most strong chaos arises for $|\phi|=\pi/2$
(solid line) corresponding to the initial values of resonant phases
$|\psi_{2,3}(0)|$ most distant from the integrable limit.
\par
\begin{figure}[htbp!]
\includegraphics[width=0.8\linewidth]{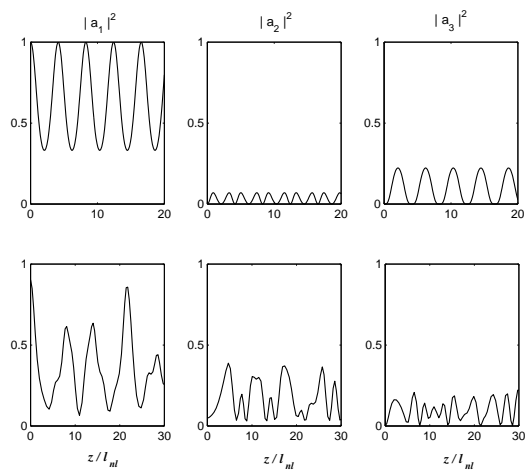}
\caption{Regular (upper) and chaotic (lower) spatial evolutions
of scaled intensities of lightwaves at the first order QPMs.
For the upper subplot $\alpha=1$ and $\phi=0$, while for the lower subplot
$\alpha=0.95$ and $\phi=\pi/2$.}
\label{figps2}
\end{figure}

The motion is always regular for the standard THG ($\alpha=1$), as well as for
some range
of $\alpha$ in the vicinity of $\alpha=1$ (see the right hand side of
Fig.~\ref{figps1}). A regular spatial evolution of lightwaves for $\alpha=1$
is shown in the upper subplot of Fig.~\ref{figps2}.
However, for $|\phi|=\pi/2$ a strong chaos exists already
for $\alpha\approx 0.95$, {\it i.e.} for $a_1(0)=0.95$, $a_2(0)\approx
0.22 i$, $a_3(0)=0$, see lower subplot of Fig.~\ref{figps2}.
Thus, the possibility of transition to chaos must be taken into account
in the application of an additional pump of frequency
$2\omega$ in order to increase an efficiency of THG \cite{egorov}.
\par
We consider now the situation corresponding to the left hand side of
Fig.~\ref{figps1} with $\alpha\ll 1$. This is the parametric amplification
with a low-frequency pump \cite{chirkin-review}. In this case our analysis
demonstrates that the evolution of waves is weakly chaotic for
$|\psi_{2,3}(0)|$ distant from the integrable limit. In this regime, the
Lyapunov exponent has some very small but yet positive value, therefore
it is very difficult to distinguish a weak chaos from a regular motion.
In practical terms, it means that one needs to have a very long sample to
see the differences between regular and weakly chaotic spatial
evolutions of light waves.
\par
\begin{figure}[htbp!]
\includegraphics[width=0.8\linewidth]{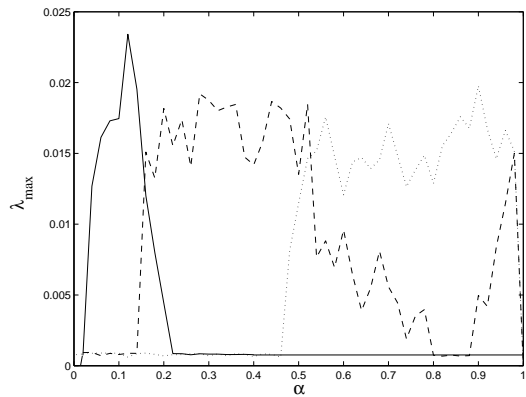}
\caption{Same as in Fig.~\ref{figps1} but for the high order QPMs (problem 1,
set II):
$\phi=-\pi/2$ (solid line),
$\phi=-0.1$ (dashed line) and $\phi=-0.01$ (dotted line).}
\label{figps3}
\end{figure}

Now we turn to the consideration of nonlinear dynamics using the second set
of QPM
parameters but same set of initial conditions (``Set II, Problem 1'').
Main results
on the transition to chaos are depicted in Fig.~\ref{figps3}.
Again, as in
Fig.~\ref{figps1}, $|\psi_2(0)|=|\psi_3(0)|=|\phi|=0$ results in a
regular motion, while a motion is chaotic for many initial conditions if
$|\phi|>0$. However, the absolute values of Lyapunov exponent are small: Really,
$\max{\lambda_{max}}\simeq 0.1$ in  Fig.~\ref{figps1}, but
$\max{\lambda_{max}}\simeq 0.01$ in  Fig.~\ref{figps3}.
Therefore, we conclude that multiple interaction of waves employing
high order QPMs is more stable against a transition to chaos in
comparison with the case of first order QPMs.
\par
\begin{figure}[htbp!]
\includegraphics[width=0.8\linewidth]{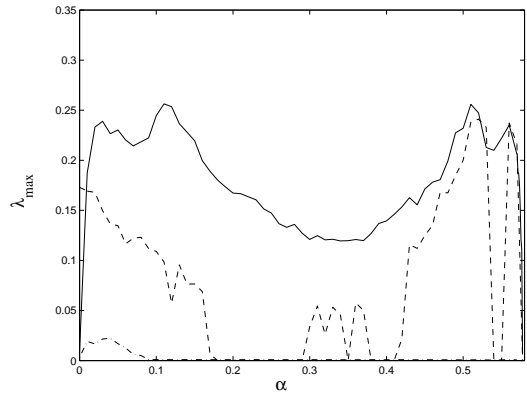}
\caption{Dependence of the value of maximal Lyapunov exponent
on the amplitude of third wave at the boundary of optical superlattice,
$\alpha$, and
for the different phases and the QPM orders (problem 2,
sets I and II): $\theta_1=\pi/2$, $\theta_2=0$,
$\theta_3=\pi$, first order QPMs (solid line);
$\theta_1=\theta_2=\theta_3=-\pi/2$, first order QPMs (dashed line);
$\theta_1=\pi/2$, $\theta_2=0$, $\theta_3=\pi$, high order QPMs
(dashed and dotted line).}
\label{figps4}
\end{figure}

We consider now a nonlinear dynamics in the case when some portion of energy
is presented at $z=0$ in each of interacting waves
(``Problem 2''). We present our findings in Fig.~\ref{figps4}.
A strong chaos arises as soon as one of
the resonant phases becomes different from the integrable limit
$|\psi_{2,3}(0)|=0$
($|\psi_{2}(0)|=\pi$ and $|\psi_{3}(0)|=\pi/2$ for a solid line,
$|\psi_{2}(0)|=\pi/2$ and $|\psi_{3}(0)|=0$ for a dashed line).
We should note that for the parameters corresponding to a solid curve in
Fig.~\ref{figps4} strong chaos exists for almost all values of initial
wave
amplitudes $\alpha$. Chaos is sufficiently weaker for the high order QPMs in
comparison with the case of first order QPMs: {\it cf} a dashed line with
a dashed and dotted line that correspond to the same values of phases
$\theta_j$ but to the different sets of QPM-parameters.
\par
Finally, we analyze the set of initial conditions termed as ``Problem 3''.
In particular, it includes the down conversion \cite{komissarova,chirkin-review}
or, in other words the fractional conversion $\omega\rightarrow (2/3)\omega$
\cite{konotop00}, in the case $\alpha\ll 1$. For this set of initial conditions
we did not find visible regions of chaotic dynamics.
\par
In order to reliably distinguish regular and chaotic spatial evolutions of
lightwaves in conditions of an experiment,
one needs to have many enough characteristic nonlinear lengths, $l_{nl}$,
on the total length of the crystal $L$: $L/l_{nl}\gtrsim 10$
\cite{alekseev,alekseeva}. Importantly, it appears possible to meet this
condition in the typical NPCs. Really, for periodically poled lithium
niobate with
a period $\Lambda=30$ $\mu$m, a crystal length $L\simeq 1$ cm, a nonlinear
coefficient $d_{33}=34$ pm/V \cite{pfister,volkov} and a light intensity
$A_0^2=0.76$ GW/cm$^2$ ($\lambda=1.064$ $\mu$m) \cite{vidakovic}, we have
$L/l_{nl}\simeq 100$.
Moreover, chaos should be easier observable in the GaAs optical
superlattice with $d_{14}\geq 90$ pm/V \cite{eyres01}.
\par
In summary, we have shown that a simultaneous multiwavelength generation in
typical nonlinear photonic crystals is often chaotic. This fact must be
taken into an account in realization of compact laser multicolor sources for
printers, scanners, and color displays based on the quasi-phase-matched
harmonics generation.
\par
We should distinguish our results from a recent paper \cite{buryak00},
where nonlinear spatial field dynamics and chaos have been studied in
a quadratic media with a periodic Bragg grating.
\par
We thank Andreas Buchleitner for discussions and support, Martin
Fejer for useful comments, and Pekka Pietil\"{a}inen for critical
reading of the manuscript. KNA is also grateful to co-authors of
his works \cite{alekseev,alekseeva} for a creative cooperation.
This research was partially supported by the Academy of Finland
(grant 100487) and the MPI PKS.

\end{document}